\documentstyle[preprint,aps]{revtex}
\begin{document}
\title{On  the  $W$-algebra  in  the   Calegero-Sutherland   model
using the Exchange operators}
\author{V. Narayanan and $^*$M. Sivakumar}
\address{School of Physics\\
University of Hyderabad\\
Hyderabad - 500 046 (INDIA)}
\maketitle
\begin{abstract}

We study the $W_\infty$ algebra in the Calegero-Sutherland  model
using the exchange operators. The presence of all the sub-algebras
of $W_\infty$  is shown in this model. A simplified proof for  this
algebra, in the symmetric ordered basics, is given. It is pointed
out  that  the  algebra  contains  in  general,  nonlinear  terms.
Possible connection to the nonlinear $W_\infty$ is discussed.
\vskip1.5in
$^*$ email: ms-sp@ uohyd.ernet.in
\end{abstract}
\newpage
In  recent   times   there   is   a   revival   of   interest   in the
Calegero-Sutherland model [1] due  to  its  relevence  in  various
fields like Quantum Hall Effect [2], Spin Chains  [3],  Fractional
Statistics [4], Quantum Chaos [5], 2D gravity [6], 2D  QCD [7].  Apart
from its application to these and  other  fields,  this  model  is
interesting in its own right due  to  its  classical  and  quantum
integrability. Also this one dimensional many-body system, is  the
only known example for which certain dynamic correlation functions
can be exactly computed [8]. Despite these remarkable developments
in recent years, the model is not completely solved, in  the  sense
that the complete excited eigenstates are  not  known  explicitly.
Another interesting, perhaps related, problem is with  regards  to
the symmetry  algebra  in  this  model.  Recently,  using  Quantum
Inverse Scattering Method (QISM) [9] and  earlier,  by  collective
field method [10], $W_\infty$ algebra [11] was shown to be present
in this model. In this paper we study the  $W_\infty$  algebra  in
this  model,  using  the  recently  developed  exchange   operator
formulation [12]. As will be shown later, this formalism is well
suited to study the $W_\infty$ algebra.

In  this  formalism,  ordinary  derivative  is  replaced   by   a
`covariant derivative' (known as Dunkl derivative [13]) with  the
gauge field part containing an exchange  operator  $M_{ij}$,  with
the following properties:
$$
M_{ij}\phi_j = \phi_i M_{ij} \qquad (i,j=1\cdots N) \eqno(1a)
$$
for any operator $A_i$ (including $M_{ij}$ itself)
$$
M_{ij}\phi_k  =  \phi_k M_{ij} \qquad (i,j\ne k)\eqno(1b)
$$
$$
M_{ij}  =  M_{ji}~~~~~~~~~~~~~~\eqno(1c)
$$
$$
M_{ij}^2 = 1 ~~~~~~~~~\eqno(1d)
$$
\addtocounter{equation}{1}
This covariant derivative, is explicitly given as $[h=m=1]$
\begin{equation}
\pi_i   =    p_i + i\lambda    \sum_j^\prime
{1\over(x_i-x_j)}  M_{ij}
\end{equation}
where $\lambda$ is a constant and prime over  the  summation  $j$,
indicates the absence of $j=i$ term. This derivative $\pi_i$,  obeys
the zero curvature condition:
\begin{equation}
[\pi_i, \pi_j] =0
\end{equation}
The Hamiltonian for the Calegero model with  the  common  harmonic
confinement (of frequency $\omega=1$), is given as
\begin{equation}
H   =   -{1\over2}   \sum_{i=1}^N   \pi_i^2+   {1\over2}    \omega^2
\sum_{i=1}^N x_i^2
\end{equation}
Note that this is diagonal in  the  particle  index  and  formally
resembles that of harmonic oscillator.  This  Hamiltonian  is  the
same as that of the Calegero Hamiltonian
\begin{equation}
H_c  =  {1\over2}  \sum_{i=1}^N   p_i^2   +   {1\over2}   \sum_{i>j}
{\lambda(\lambda-1)\over(x_i-x_j)^2}  +  {1\over2}  \omega^2  \sum_{i=1}^N
x_i^2
\end{equation}
for the physical states obeying the Bosonic  condition,
\begin{equation}
M_{ij}|\psi> = + |\psi>
\end{equation}
Due to the formal similarity of the Hamiltonian (4) with  that  of
the harmonic oscillators, define a  generalised  annihilation  and
creation operators $(a_i, a_i^\dagger)$ as
\begin{eqnarray}
a_i & = & {\pi_i-ix_i\over\sqrt{2}} \nonumber\\
a^\dagger_i & = & {\pi_i+ix_i\over\sqrt{2}}
\end{eqnarray}
using which (4) can be expressed as
\begin{equation}
H = {1\over2} \sum_{i=1}^N (a_ia^\dagger_i + a^\dagger_i a_i)
\end{equation}
The commutation relation between $a_i$ and $a_i^\dagger$ can easily
shown to be
\begin{eqnarray}
[a_i, a_j^\dagger] &  =  &  \delta_{ij}(1+\lambda  \sum_k^\prime
M_{ik}) - (1-\delta_{ij}) \lambda M_{ij}\\
{}[a_i, a_j] & = & [a_i^\dagger, a_j^\dagger] = 0
\end{eqnarray}
Also, it can be shown,
\begin{equation}
[H, a_i (a_i^\dagger)] = -a_i(+a_i^\dagger)
\end{equation}
similar to that of harmonic oscillator.

The problem we address here, is about  the  $W$  algebra  in  this
model using such a  formulation.  The  exchange  operator  method,
discussed above, is well suited  for  this  purpose,  due  to  its
formal resemblence to the oscillator problem. It is known that the
generators of $W_\infty$ algebra  can be  expressed  in  terms  of
the oscillator algebra. In one of  the basis [14],
$$
W_{nm} = a_o^nA^m\,\, \qquad n,m\ge 1
$$
where $A\equiv  a_oa_o^\dagger$,  and  $a_o,a_o^\dagger$  satisfies
$[a_o, a_o^\dagger] =1$.

The commutation relation of the differential operators, is
\begin{equation}
[a_o^nf(A),   a_o^mg(A)]   =    a_o^{n+m}\left(f(A-m)    g(A)    -
f(A)g(A-n)\right)
\end{equation}
where $f$ and $g$ are polynomials.

This has as the sub-algebra,
\begin{itemize}
\item[(a)] (centreless) Virasoro algebra generated by $L_n=-a_o^nA$
$$
[L_n,L_m] = (n-m)L_{nm}\eqno(13a)
$$
\item[(b)] $[L_o, W_{nm}] = -n W_{nm}$ \hfill(13b)
\item[(c)] $[W_{ok}, W_{ol}]=0$  (Cartan-sub algebra) \hfill (13c)
\end{itemize}
One can also, consider the basis [15], in which
$$
W_n^{(s)} \equiv a_o^{s-n-1} a_o^{\dagger\, s+n-1} \eqno(14a)
$$
obeys the algebra
$$
[W_n^{(s)}, W_m^{(t)}] = [n(t-1)-m(s-1)] W_{n+m}^{s+t-2} +  \cdots
 \eqno(14b)
$$
\addtocounter{equation}{2}
where $\cdots$ denotes the  lower  order  terms,  due  to  quantum
correction, and its coefficients depend on  the  ordering  in  the
definition of $W_n^{(s)}$.

Due  to  the  realisation  of  $W_\infty$,  algebra  in  terms  of
oscillator algebra, it is pertinent  to  ask,  if  the  oscillator
algebra is replaced by the modified one given  by  (10),  does  it
give rise to the same  $W$  algebra?  This  is  the  question,  we
attempt, in this paper, Bergshoeff and Vasilev [16], have shown Virasaro
algebra, in the same formulation. The quantum integrability of this
model, shown by Polychronokos [12], implies the  presence  of  the
Cartan  sub-algebra  (13c).  Ujino  and  Wadati [9],  have  used   the
Lax operators to show $W_\infty$ algebra, after  long  and  tedious
calculations. In contrast, we  find  that  the  exchange  operator
formalism simplifies the proof enormously. Also, we show that,  in
contrast to Ujino and Wadati [9], there are non-linear terms  in
the algebra. This paper is organised as follows:  Section  I  shows  the
sub-algebra's mentioned in (13) using the  oscillator  defined  in
(10). Section II, deals with the symmetric  ordered  form  of  the
basis in (14a). Finally, we end with the conclusion.

\noindent{\bf Section I:}

In this we show the presence of the sub-algebras of $W_\infty$  in
the basis given in (13). The presence of Virasavo algebra, in  the
basis (13), was shown by Bergshoeff and Vasiliev [16]. Due to  the
restriction $n,m>0$, central term is  absent.  Next  we  show  the
sub-algebra   (13b)    with    $a_o,a_o^\dagger$    replaced    by
$a_i,a_i^\dagger$ obeying (9),
\begin{eqnarray}
{}[L_o, W_{nm}] & = & \left[\sum_{i=1}^{N^\prime} A_i,  \sum_{j=1}^N
a_j^nA_j^m\right] \\
& =  &  \sum_{i=1}^N\left[A_i,  a_i^n\right]A_i^m  +  \sum_{i,j}^\prime
[A_i, a_j^n]A_j^m + \sum_{i,j}^\prime a_j^n[A_i,A_j^m]
\end{eqnarray}
Here the prime over the summation implies $i=j$ term is excluded.
Using the commutation rules (9) and (10) and the Leibnitz rule, we
get

\begin{eqnarray}
&  =  &  -n\sum_{i=1}^N  a_i^n  A_i^m   -   \lambda   \sum_{i=1}^N
\sum_{\beta=o}^{n-1}    a_i^{\beta+1}    M_{ij}    a_i^{n-1-\beta}
A_i^m\nonumber\\
&  +  &  \lambda a_i  \sum_{\beta=o}^{n-1}  a_j^\beta
M_{ij}   a_j^{n-1-\beta}   A_j^m    +    \lambda\sum_{ij}    a_j^n
\sum_{\beta=o}^{m-1} A_j^\beta(A_i-A_j) M_{ij}A_j^{m-1-\beta}
\end{eqnarray}
The last term in the above equation follows from using,
\begin{equation}
[A_i, A_j] = \lambda(A_i-A_j) M_{ij}
\end{equation}
which follows from using (9) and (10) in the definition of  $A_i$.
Next we show that all terms dependent  on  $\lambda$  cancel  with
each  other,  leading  to  the  desired  result.  Collecting   only
$\lambda$ dependent terms, seperating the sum over $\beta$ in  the
second term (and the third term) as  $\beta=n-1$  ($\beta=0$)  and
the rest, and also using the property (2.1) of $M_{ij}$, we get
\begin{eqnarray}
&&\lambda   \sum_{ij}^\prime\biggl[-a_i^nA_j^m   -    \sum_{\beta=o}^{n-2}
a_i^{1+\beta}    a_j^{n-1-\beta}    A_j^m    +    a_i^nA_i^m     +
\sum_{\beta=1}^{n-1}a_i^{n-\beta}a_j^\beta A_i^m\nonumber\\
& + &  \sum_{\beta=o}^{m-1}a_j^nA_j^\beta A_i^{m-\beta}  -  a_j^n
\sum_{\beta=o}^{m-1}
A_j^{\beta+1}A_i^{m-1-\beta}\biggr]M_{ij}
\end{eqnarray}
Note that $\beta=0$ ($\beta=m-1$) in the fifth term  (sixth  term)
cancels with the first (third) term in (19).  Upon  interchanging
$i,j$ in the second term, in (19), we find all of  them  cancels.
More explicitly,
\begin{eqnarray}
\lambda\sum_{ij}^\prime  && \biggl[\left(-\sum_{\beta=1}^{n-1}
a_i^\beta                     a_j^{n-\beta}                      +
\sum_{\beta=1}^{n-1}a_j^{n-\beta}a_i^\beta\right)A_j^m\nonumber\\
&  +  &  a_j^m\left(\sum_{\beta=1}^{m-1}  A_j^\beta  A_i^{m-\beta}
-\sum_{\beta=1}^{m-1}  A_j^\beta   A_i^{m-\beta}\right)
\biggr] M_{ij} = 0
\end{eqnarray}
Thus we get
$$
[L_o, W_{nm}] = -n W_{nm}
$$
The Cartan sub-algebra follows from the quantum  integrability  of
the     model.     It     was     shown      by      Polychronokos,
$\displaystyle{\sum_{i=1}^N A_l^n\equiv I_n}$, obey $[I_n,I_m]=0$.
It is interesting to note that all the sub-algebras' are satisfied
as an operator relation, without having to use the physical  state
condition (6), for only when the latter holds, this is  equivalent
to Calegero-Sutherland model.
\vskip0.5cm
\noindent{\bf Section II}

In order to  study  the  full  $W_\infty$  algebra,  it  is  found
convenient  to  consider  the  basis,  in  which  generator  is  a
symmetric combination of arbitrary powers of $a^n$ and $a^{\dagger
m}$.
\begin{equation}
O_{nm}   \equiv   {n!m!\over(n+m)!}    \sum_{i=1}^N    \left(a_i^n
a_i^{\dagger  m}  +  a_i^{n-1}a_i^{\dagger  m}  a_i  +  \cdots   +
a_i^{\dagger m} a_i^n\right)
\end{equation}
This is the basis used by Ujino and Wadati to show  $W_\infty$  in
Calegero mode. Note that the normalisation $(n+m)!/n!m!$ gives the
number of terms in the symmetric combination.

First we show that
\begin{equation}
\left[O_{nm}  ,  a_i  (a_i^\dagger)  \right]  =  -m   O_{n,m-1}(i)
(nO_{n-1,m}(i))
\end{equation}

A similar equation with $a_i$, and $a_i^\dagger$ replaced  by  the
Lax operators, called as a `generalised Lax equation' was  derived
by Ujino and Wadati after a lengthy algebra, and this  formed  the
basis for their proof of $W_\infty$ algebra in this model. We show
that, in contrast, this can be arrived with great ease, in the exchange
operator formulation.

First note that the symmetric combination can be generated by  the
generating function
\begin{equation}
D(i)  \equiv  \alpha a_i + \beta a_i^\dagger
\end{equation}
by defining $O_{nm}$ to be
\begin{equation}
O_{nm}  =    {1\over(n+m)!}  \partial_\alpha^n  \partial_\beta^m
\sum_{j=1}^N D^{n+m}(j)\biggr|_{\alpha=\beta=0}
\end{equation}
As an example,
\begin{equation}
O_{23}    =    {1\over    5!}    \partial_\alpha^2\partial_\beta^3
\sum_{j=1}^N    (\alpha     a_j     +     \beta     a_j^\dagger)^5
\biggr|_{\alpha=\beta=0}
\end{equation}

Due to the  condition  $\alpha=\beta=0$,  only  those  terms  with
coefficient $\alpha^2\beta^3$ contribute,
\begin{eqnarray}
&     =     &     \partial_\alpha^2\alpha^2\partial_\beta^3\beta^3
\biggr|_{\alpha=\beta=0}  \sum_j  \left(a_j^2a_j^{\dagger  3}+\cdots  +
\mbox{symmetric combination}\right)\nonumber\\
O_{23} & = & {2!3!\over5!} \sum_{i=1}^N\left(a_i^2a_i^{\dagger3}  +
\cdots + a_i^{\dagger3}a_i^2\right)\nonumber\\
{}[O_{nm},a_i] & = & {1\over(n+m)!} \partial_\alpha^n\partial_\beta^m
\left[\sum_{j=1}^N D^{n+m}(j),  a_i\right]\biggr|_{\alpha=\beta=0}
\\
& = & {1\over(n+m)!} \partial_\alpha^n\partial_\beta^m\sum_{j=1}^N
\sum_{l=o}^{n+m-1}      D^l(j)      [D(j),      a_i]D^{n+m-l-1}(j)
\biggr|_{\alpha=\beta=0}
\end{eqnarray}

Using (9) and (10), with (23) inserted in (24), we get
\begin{eqnarray}
&    =    &    {1\over(n+m)!}    \partial_\alpha^n\partial_\beta^m
\sum_{l=o}^{n+m-1}\biggl[D^l(i)\left(-\beta-\lambda          \beta
\sum_{j\ne i} M_{ij}\right) D^{n+m-l-1}(i)\nonumber\\
& + & \sum_{j\ne i} D^l(j)  \beta  \lambda  M_{ij}D^{n+m-l-1}  (j)
\biggr]_{\alpha=\beta=0}\nonumber\\
& =& {1\over(n+m)!} \partial_\alpha^n\partial^m_\beta \biggl|-\beta
(n+m)D^{n+m-1}(i)  -  \lambda  \sum_{j\ne  i}   \sum_{l=o}^{n+m-1}
\nonumber\\
&\times       &        \left(D^l(i)        D^{n+m-l-1}(j)        -
D^l(j)D^{n+m-l-1}(i)\right)M_{ij}\biggr]_{\alpha=\beta=0}
\end{eqnarray}
Note that the $\lambda$ dependent terms can be expressed as
$$
\lambda     \sum_{j\ne      i}      \sum_{l=o}^{n+m-1}      \left[
D^l(i)~,~D(j)^{n+m-l-1}\right]
$$
which vanishes on using
$$
\left[ D(i)~,~D(j)\right] = 0
$$
Thus we get
\begin{eqnarray}
{}[O_{nm},a_i]        &        =         &         -{1\over(n+m-1)!}
\partial_\alpha^n\partial_\beta^m(\alpha^n                \beta^m)
\left[a_i^na_i^{\dagger      m-1}+\cdots      +       a_i^{\dagger
m-1}a_i^n\right]\nonumber\\
& = & -m O_{n,m-1}(i)
\end{eqnarray}
Similarly, it can be shown that
$$
[O_{nm}, a_i^\dagger] = nO_{n-1,m}(i)
$$
Similar relation with  $a_i,  a_i^\dagger$  replaced  by  the  Lax
operators, was obtained by Ujino and Wadati [9]. In contrast  to
their approach, here the similar result  has  been  obtained  with
great ease.
\begin{eqnarray*}
W^{(s)}_n & \equiv & O_{s-n-1,s+n-1}\\
{}[W_n^{(s)},     W^{(t)}_m]     &     =     &      \left[W_n^{(s)},
{(s-n-1)!(s+n-1)!\over(2s)!}  \sum_{i=1}^N   a_i^{\dagger   s-n-1}
a_i^{s+n-1} + \cdots \right]
\end{eqnarray*}
Using (22),
\begin{equation}
[W_n^{(s)}, W_m^{(t)}]  =  2(n(t-1)  -  m(s-1))W_{n+m}^{(s+t-2)}  +
\cdots
\end{equation}
This  defines  the  $W_\infty$  algebra.  The  dotted  terms,  are
obtained when  the  commutation  relation  (9)  and  (10) are used
to express the leading term in (30) as $W_\infty$ generator with that
structure constant.  In
general, there will  be  terms  (apart  from  linear  lower  order
$\lambda$ independent terms) which are non-diagonal in $i$ and $j$.
These, in general, will  be  $\lambda$  dependent,  with  $M_{ij}$
relating the $i$ and $j$ indices, and the summation  over  all
$i,j$ with $i\ne j$. Explicitly we will have a generic form,
\begin{equation}
\lambda^p  \sum_i  (a_i^{n_i}a_i^{\dagger  m_i})   \sum_{j\ne   i}
(a_j^\dagger)^{n_2}(a_j)^{m_2} M_{ij}
\end{equation}
with $p\ge 1$, $n_i, m_i (i=1,2)\ge 0$. Since  the  equivalence  of
the Hamiltonian (4) to  the  usual  Calegero-Sutherland  model  is
valid only on the physical states, obeying the condition  (6),  on
a  physical  states,  $M_{ij}$  drops  out (This some times leads to
$N$ dependence).  Then  in  the   terms
$\sum_{j\ne i} a_j^{\dagger n_2} a_j^{m_2}$, replacing it with
$$
\sum_{j=1}^N\left(a_j^{\dagger       n_2}       a_j^{m_2}        -
(a_i^\dagger)^{n_2}(a_i)^{m_2}\right),
$$
the former appears  to  be  non-linear  in  the  generator  (after
symmetrization). The latter is diagonal  in  the  particle  index.
Thus in general a $\lambda$-dependent non-linear algebra  results.
But the linear term also can have $\lambda$ dependent coefficient.
Even in the QISM [9], such non-linear terms  in  the  $W$  algebra
are, in general, possible, although, it was not explicitly  stated
in their work.

As an example, $[W_{-1/2}^{5/2}, W^{5/2}_{1/2}]$ is calculated, using
the Eqns. (9,10) and (29) in the above basis,
\begin{eqnarray}
[W_{1/2},^{5/2}, W^{5/2}_{-1/2}] & = & -3W_o^3 -  {1\over2}  \lambda^2
\sum_{ij}^\prime M_{ij}^2 + {1\over2}\sum_i[a_i, a^\dagger_i]\nonumber\\
& = & -3W_o^3 - {1\over2} \lambda^2N(N-1) + {1\over2}(N+  \lambda
\sum_{ij}^\prime M_{ij}|)
\end{eqnarray}

Acting on the physical  states  due  to  (9),  the  last  give
$N(N-1)$
$$
= -3W_o^3 + {1\over2} N + \lambda {(1-\lambda)\over2} N(N-1)
$$
Since
$$
\sum_{i=1}^N O_{oo} = W_o^1 = N
$$
we get
\begin{equation}
[W^{5/2}_{1/2}, W^{5/2}_{-1/2}] = - 3W_o^3  +  {1\over2}  W_o^1  +
{\lambda(1-\lambda)\over 2} W_o^\prime(W_o^\prime-1)
\end{equation}
Thus there are nonlinear terms also.

Similarly another example is given  in  the  basis,  generated  by
$a_i$, and  $A_i$,  discussed  earlier.  With  the  definition  of
$\displaystyle{W_{nm}=\sum_{i=1}^N   a_i^nA_i^m}$   consider   the
commutator $[W_{11},W_{12}]$. After straightforward algebra, using
(9,10) and (18), we get
\begin{eqnarray*}
[W_{11}, W_{12}] & = & W_{22}-W_{21} -  2\lambda\sum_{ij}^\prime  a_i^2
A_j M_{ij}\\
& + &  \lambda  \sum_{ij}^\prime  a_j^2  A_j  M_{ij}  +  \lambda^2
\sum_{i\ne j \ne k} a_j^2 A_k M_{jk}M_{ij} - \lambda^2 \sum_{i\ne j
\ne k} a_i^2A_jM_{ik}M_{ij}
\end{eqnarray*}
Using the physical state condition (2.14a), terms with $\lambda^2$
as coefficient cancels, and the rest  are given by,
\begin{equation}
= W_{22} - W_{21} - 2\lambda\sum_{ij}^\prime  a_i^2  A_j  +\lambda
(N-1)\sum_{ij}^\prime a_j^2 A_j
\end{equation}
In the third term which is non-diagonal in $i$ and $j$, express it
as
\begin{eqnarray*}
&=& 2\lambda \sum_i a_i^2\left(\sum_{j\ne i} A_j\right)\\
& = & 2\lambda \sum_i a_i^2\left(\sum_jA_j - A_i\right)\\
& = & 2\lambda(W_{20} W_{01} - W_{21})
\end{eqnarray*}
Using $N=W_{00}$, the final commutator is
\begin{eqnarray}
[W_{11}, W_{12}] & = &  W_{22}-W_{21}-2\lambda(W_{20}W_{01}-W_{21})  +
\lambda (W_{00}W_{21}-W_{21})\nonumber\\
& = & W_{22}-W_{21} - 2\lambda W_{20} W_{01} +  \lambda  W_{00}W_{21}  +
\lambda W_{21}
\end{eqnarray}
Thus, this also displays non-linear terms in the algebra.

\noindent{\bf Discussion:}

In this paper, we have studied the symmetry algebra in the quantum
Calegero-Sutherland model, using the exchange operator  formalism.
In this, operators $a_i, a_i^\dagger$ are  introduced  (8),  which
obey algebra like that  of  oscillator  (11).  Since  $W$  algebra
generators can be expressed in terms of ordinary $(\lambda\to  0$
case of (9)), oscillator we  ask  about  the  $W_\infty$  algebra,
involving the modified oscillator (7), obeying the  relation  (9).
We showed that the sub-algebras of $W_\infty$ are  obeyed,  as  an
operator relation. We also gave a  simpler  proof,  compared  with
Ref.(9), for $W$ algebra in this  model.  But  in  contrast  to  them,
existence of terms non-linear in the generator, in the  commutation
algebra is pointed out, when the physical state condition  is  (6)
imposed. The non-linear term are always  $\lambda$  dependent.  It
should be interesting to see if the $W$ algebra present here is
isomorphic to the non-linear $\hat{W}_\infty$ algebra proposed by
Yu and Wu [17].
The latter is also centreless and is a two-parameter deformation
of $W_\infty$. One parameter signifies the non-leading terms and
the other non-linear terms. The Calegero model has two parameters
$\lambda$ and $\hbar$ (when the latter is restored), but appearing
in the combination $\lambda \hbar$. It remains to be seen, if by
suitable scaling and taking appropriate linear combination of its
generators, a connection between these two algebras can be made.

Such  a  connection,  will  be  of
interest, as the $\hat{W}_\infty$ is known  to  provide  the  second
Hamiltonian
structure to $KP$ hierarchy [17]. Such  a  connection  should  also  be
useful to prove integrability of the continuum Calegero-Sutherland
model [18]. This may also shed light on the relationship between
Calegero-Sutherland model and two-dimensional physics.
 Apart from these, if should be possible to extend the  results
of this work to spin generalization  of  Calegero  model,  and  to
lattice integrable systems [19].

\noindent{\bf Note  added:}  When  this  work  was  completed,  we
received the ref. [20], which also points out the non-linear terms
in $W$-algebra in the Calegero model. We thank Ujino, for
correspondance and  for sending us the ref. (20).
\vskip0.5cm
\noindent{\bf Acknowledgement:}

M.S. thanks M.V.N. Murthy and R. Shankar for initial collobaration
and Institute of Mathematical Sciences, for hospitality where  this
work was started. We  thank  Prashanta  K.  Panigrahi  for  useful
comments on the manuscript.
\newpage
\noindent{\bf References}
\begin{enumerate}
\item F. Calogero,  J.  Math.  Phy7s.  {\bf10},  2191  (1969);  B.
      Sutherland, J. Math. Phys. {\bf12}, 246 (1971);  Phys.  Rev.
      A{\bf4}, 2019 (1971).
\item Azuma and S. Iso, Phys.  Lett.  B{\bf331},   (1994) 107;  N.
      Kawakami, Phys. Rev. Lett. {\bf71}, 275 (1993); P.K. Panigrahi
	and M. Sivakumar, Phys. Rev. B. (in press).
\item F.D.M. Haldane, Phys. Rev. Lett. {\bf60}, 635  (1988);  B.S.
      Shastry, Phys. Rev. Lett. {\bf60}, 639 (1988).
\item A.P.  Polychronakos,  Nucl.  Phys.  B{\bf324},  597  (1989);
      M.V.N. Murthy and R.  Shankar, Phys.  Rev.  Lett. {\bf73}, 3331 (1994).
\item B.D. Simons, P.A. lee and B.L. Altshuler, Phys.  Rev.  Lett.
      {\bf70}, (1993) 4122.
\item I. Andric, A. Jevicki and H. Levine, Nucl.  Phys.  B{\bf215}
      (1983) 307; A. Jevicki, Nucl. Phys.  B{\bf376}  (1992)  75.;
\item J.A. Minahan and A.P. Polychronakos, Phys.  Lett.  B{\bf312}
      (1993) 155; ibid {\bf336} (1994) 288.
\item Z.N.C. Ha, Phys. Rev. Lett. {\bf73}, 1574 (1994).
\item H. Ujino and M. Wadati, J. Phys. Soc.  Jpn.  {\bf63}  (1994)
      3585; K. Hikami and M. Wadati, J. Phys. Soc. Jpn.  {\bf  62}
      (1993) 4203; Phys. Rev. Lett. {\bf73} (1994) 1191.
\item J. Avan and A. Jevicki, Phys.  Lett.  B{\bf272}  (1991)  17;
      Commun. Math. Phys. {\bf150}, 149 (1992).
\item P. Bowknegt  and  K.  Schoutens,  Phys.  Rep.  {\bf223}  183
      (1993).
\item A.P. Polychronakos, Phys. Rev. Lett. {\bf69} (1992) 703;  L.
      Brink,  T.H.  Hansson  and  M.A.   Vasiliev,   Phys.   Lett.
      B{\bf286} (1992) 109.
\item C.F. Dunkl, Trans A.M.S. {\bf311} (1989) 167.
\item H. Awata, M. Kukuma, Y. Matsuo and S. Odake - hepth/9408158,
      Talk at Quantum field theory, integrable models  and  beyond,
      YITP.
\item A. Capelli, C.A. Trugenberger and  G.R.  Zemba,  Phys.  Rev.
      Lett. {\bf72} (1994) 1902.
\item E. Bergshoeff and M.A. Vasiliev, Int. J. Mod. Phys. A{\bf 10}
	3477 (1995).
\item F. Yu and Y.S. Wu, Nucl. Phys. B{\bf373} (1992) 713.
\item A.P. Polychronokos, Phys. Rev. Lett. {\bf74}, 5153 (1995).
\item J.A. Minahan and A.P. Polychronokas, Phys.  Lett.  B{\bf302}
      (1993) 265.
\item H. Ujino and M. Wadati, J. Phys. Soc.  Jpn.  {\bf64}  (1995)
      39.
\end{enumerate}

\end{document}